\documentclass[twocolumn,showpacs,preprintnumbers,amsmath,amssymb,superscriptaddress]{revtex4}

\usepackage{graphicx}
\usepackage{dcolumn}
\usepackage{bm}

\begin{document}


\title{Nonlinear coupling of continuous variables at the single quantum level}
\author{C. F. Roos}
\email{christian.roos@uibk.ac.at}
\affiliation{%
Institut f\"ur Experimentalphysik, Universit\"{a}t Innsbruck,
Technikerstra{\ss}e 25, A--6020
Innsbruck, Austria\\
}%
\affiliation{%
Institut f\"ur Quantenoptik und Quanteninformation der
\"Osterreichischen Akademie der
Wissenschaften, Technikerstra{\ss}e 21a, A--6020 Innsbruck, Austria\\
}%
\author{T. Monz}
\affiliation{%
Institut f\"ur Experimentalphysik, Universit\"{a}t Innsbruck,
Technikerstra{\ss}e 25, A--6020
Innsbruck, Austria\\
}%
\author{K. Kim}
\affiliation{%
Institut f\"ur Experimentalphysik, Universit\"{a}t Innsbruck,
Technikerstra{\ss}e 25, A--6020
Innsbruck, Austria\\
}%
\author{M.~Riebe}
\affiliation{%
Institut f\"ur Experimentalphysik, Universit\"{a}t Innsbruck,
Technikerstra{\ss}e 25, A--6020
Innsbruck, Austria\\
}%
\author{H.~H\"{a}ffner}
\affiliation{%
Institut f\"ur Experimentalphysik, Universit\"{a}t Innsbruck,
Technikerstra{\ss}e 25, A--6020
Innsbruck, Austria\\
}%
\affiliation{%
Institut f\"ur Quantenoptik und Quanteninformation der
\"Osterreichischen Akademie der
Wissenschaften, Technikerstra{\ss}e 21a, A--6020 Innsbruck, Austria\\
}%
\author{D.~F.~V.~James}
\affiliation{%
Department of Physics, University of Toronto, 60 St. George St.,
Toronto, Canada\\
}%
\author{R.~Blatt}
\affiliation{%
Institut f\"ur Experimentalphysik, Universit\"{a}t Innsbruck,
Technikerstra{\ss}e 25, A--6020
Innsbruck, Austria\\
}%
\affiliation{%
Institut f\"ur Quantenoptik und Quanteninformation der
\"Osterreichischen Akademie der
Wissenschaften, Technikerstra{\ss}e 21a, A--6020 Innsbruck, Austria\\
}%


\date{\today}

\begin{abstract}
We experimentally investigate nonlinear couplings between
vibrational modes of strings of cold ions stored in linear ion
traps. The nonlinearity is caused by the ions' Coulomb interaction
and gives rise to a Kerr-type interaction Hamiltonian
$H=\hbar\chi\hat{n}_r\hat{n}_s$, where $\hat{n}_r,\hat{n}_s$ are
phonon number operators of two interacting vibrational modes. We
precisely measure the resulting oscillation frequency shift and
observe a collapse and revival of the contrast in a Ramsey
experiment. Implications for ion trap experiments aiming at
high-fidelity quantum gate operations are discussed.
\end{abstract}

\maketitle

Kerr nonlinearities in optical media caused by dispersively acting
third-order susceptibilities play an important role in nonlinear
optics. There is considerable interest in realizing a Hamiltonian
$H=\hbar\chi\hat{n}_r\hat{n}_s$ coupling two field modes by a
cross-Kerr nonlinearity as this interaction might be used for
quantum-nondemolition measurements of single photons
\cite{Imoto85}, for the realization of quantum gate operations in
photonic quantum computation
\cite{Milburn89,Chuang95,Lloyd99,Munro05} and for photonic Bell
state detection \cite{Barrett05}. Unfortunately, the required
strong nonlinearities are difficult to realize experimentally
\cite{Turchette95,Schmidt96}.

In trapped ion quantum computing, continuous quantum variables
occur in the description of the joint vibrational modes of the ion
string. The normal modes are of vital importance for all
entangling quantum gates as they can give rise to effective
spin-spin couplings in laser-ion interactions \cite{Leibfried03}.
The normal mode picture naturally appears when the ion trap
potential is modelled as a harmonic (pseudo-)potential and the
mutual Coulomb interaction between the ions is linearized around
the ions' equilibrium positions \cite{James98}. In this way, the
collective ion motion is described by a set of independent
harmonic oscillators with characteristic normal mode frequencies.
Small deviations from this picture arise because of the
nonlinearity of the Coulomb interaction giving rise to a
cross-coupling between the normal modes. While nonlinear Coulomb
couplings between ions found lots of attention in ion trap
experiments investigating order-chaos transitions
\cite{Hoffnagle88,Bluemel88}, its effect in the normal mode regime
has mostly gone unnoticed. Here, we are not interested in resonant
mode coupling \cite{Marquet03} but rather in dispersive cross-Kerr
effects leading to shifts of the normal mode frequencies. In this
paper, we show that the Coulomb nonlinearity gives rise to a
Kerr-type Hamiltonian and we quantitatively measure the strength
of the induced frequency shift. This frequency shift is of
importance for ion trap experiments aiming at pushing the fidelity
of entangling quantum gates towards the fault-tolerant threshold.

For two cold ions of mass $m$ and charge $e$ held in an
anisotropic harmonic potential characterized by trap frequencies
$\omega_z,\omega_x=\omega_y\equiv\omega_\perp$, with
$\omega_z<\omega_\perp$, the equilibrium positions of the ions are
given by $\mathbf{r_i}=(0,0,\pm z_0)$, where
$z_0=(e^2/(16\pi\epsilon_0m\omega_z^2))^{1/3}$. Separating the
center-of-mass (COM) and the relative ion motion by introducing
the coordinates $\mathbf{R}=(\mathbf{r}_2+\mathbf{r}_1)/2$ and
$\mathbf{r}=(\mathbf{r}_2-\mathbf{r}_1)/2$ yields the potential
energy
\begin{equation}
U_{pot}=U_{pot}^{COM}(\mathbf{R})
+m{\omega_z}^2(z^2+\frac{\omega_\perp^2}{\omega_z^2}(x^2\!+y^2)+\frac{2z_0^3}{r}).
\nonumber
\end{equation}
Expanding the Coulomb potential to fourth order around $(0,0,z_0)$
by setting $z=z_0+u$ and keeping only those terms giving rise to a
cross-coupling between normal modes, we find
\begin{equation}
U_{pot}^{rel} = m{\omega_s}^2u^2+m{\omega_r}^2(x^2\!+y^2) +
V^{(3)}+V^{(4)},\nonumber
\end{equation}
where $\omega_s=\sqrt{3}\,\omega_z$,
$\omega_r=\sqrt{\omega_\perp^2-\omega_z^2}$ are the stretch and
the rocking mode frequencies \cite{King98} and where
$V^{(3)}$,$V^{(4)}$ defined by
\begin{equation}
V^{(3)}=\frac{m{\omega_s}^{\!2}}{z_0}u(x^2\!+y^2),\;V^{(4)}=-\frac{2m{\omega_s}^{\!2}}{z_0^2}u^2(x^2\!+y^2)\nonumber
\end{equation}
describe the cubic and quartic nonlinearities. Next, we use
perturbation theory for a calculation of the energy shifts
$\epsilon(n_s,n_r^{x},n_r^{y})$ of the quantum states described by
the quantum numbers $(n_s,n_r^{x},n_r^{y})$ where $n_s$ and
$n_r^{x,y}$ specify the number of stretch mode and rocking mode
phonons. In first-order, we calculate only the shift induced by
the quartic term as the cubic term does not contribute. In second
order, the quartic term can be neglected. The shift of the stretch
mode frequency by the rocking mode quantum numbers is given by
\begin{equation}
\delta\omega_s
=\frac{\epsilon(n_s+1,n_r^{x},n_r^{y})-\epsilon(n_s,n_r^{x},n_r^{y})}{\hbar}
=\chi(n_r^{x}\!+n_r^{y}\!+1)\nonumber
\end{equation}
with the cross-Kerr coupling constant
\begin{equation}
\chi=\!-\omega_s\!\left(1+\frac{\omega_s^2/2}{4\omega_r^2-\omega_s^2}\right)\left(\frac{\omega_z}{\omega_r}\right)
\left(\frac{2\hbar\omega_z}{\alpha^2mc^2}\right)^{\frac{1}{3}},
\label{freqshift}
\end{equation}
where $\alpha$ denotes the fine structure constant and $c$ the
speed of light. This expansion is valid as long as
$\omega_r/\omega_s\gg (x_0/z_0)$ and
$|2\omega_r-\omega_s|/\omega_s\gg (u_0/z_0)$ where $x_0$ ($u_0$)
are the spatial half width of the rocking (stretch) mode ground
states, respectively. The resonance $2\omega_r=\omega_s$
corresponds to a parametric coupling converting two rocking mode
phonons into a single stretch mode excitation. In experiments with
thermally occupied rocking modes, the cross-mode coupling causes
dephasing of the stretch mode motion.

In our experiments, two $^{40}$Ca$^+$ ions are confined in a
linear Paul trap with radial trap frequencies of about
$\omega_\perp/2\pi=4$~MHz. By varying the trap's tip voltages from
500 to 2000 V, the axial center-of-mass frequency $\omega_z$ is
changed from 860~kHz to 1720~kHz. The ions are Doppler-cooled on
the $S_{1/2}\leftrightarrow P_{1/2}$ transition. Sideband cooling
on the $S_{1/2}\leftrightarrow D_{5/2}$ quadrupole transition
\cite{Roos99} prepares the stretch mode in the motional ground
state $|0\rangle_s$. Simultaneous cooling of stretch and rocking
modes is accomplished by alternating the frequency of the cooling
laser exciting the quadrupole transition between the different red
motional sidebands. Motional quantum states are coherently coupled
by a laser pulse sequence exciting a single ion on the
$|S\rangle\equiv S_{1/2}(m=-1/2)\leftrightarrow|D\rangle\equiv
D_{5/2}(m=-1/2)$ transition with a focused laser beam on the
carrier and the blue sideband. Internal state superpositions
$(|S\rangle+e^{i\phi}|D\rangle)|0\rangle$ can be mapped to
motional superpositions $|D\rangle(|0\rangle+e^{i\phi}|1\rangle)$
by a $\pi$ pulse on the blue motional sideband and vice versa. We
discriminate between the quantum states $S_{1/2}$ and $D_{5/2}$ by
scattering light on the $S_{1/2}\leftrightarrow P_{1/2}$ dipole
transition and detecting the presence or absence of resonance
fluorescence of the individual ions with a CCD-camera. A more
detailed account of the experimental setup is given in Ref.
\cite{Schmidtkaler03,Schmidtkaler03b}.

\begin{figure}[t]
\includegraphics[width=8cm]{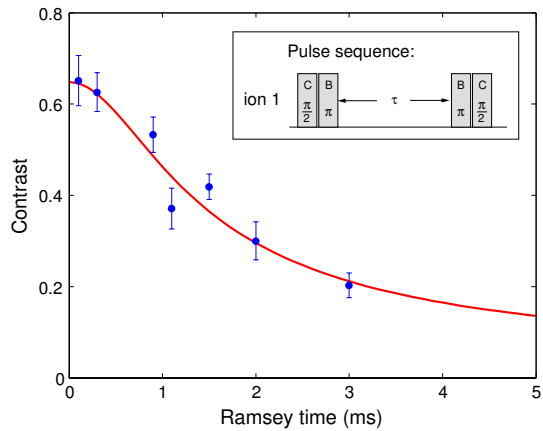}
\caption{\label{fig:RamseyContrast} Contrast $C(\tau)$ of a Ramsey
experiment on the stretch mode as a function of the Ramsey waiting
time $\tau$. The axial trap frequency was $\omega_z = (2\pi)
1486$~kHz, the radial modes were cooled close to the Doppler
limit. The initial contrast is limited by imperfect sideband
cooling, spurious excitation of the second ion by the focussed
laser beam and magnetic field noise. The subsequent loss of
contrast is caused by a dephasing of the stretch mode oscillation.
The inset shows the pulse sequence of the Ramsey experiment with
$C$ denoting carrier pulses and $B$ blue sideband pulses.}
\end{figure}
For a measurement of the stretch mode coherence, a Ramsey
experiment between motional states is performed by a laser
interacting with only one of the ions, the second ion being just a
spectator that modifies the normal mode structure. Starting from
the state $|S\rangle|0\rangle$, the superposition state
$|D\rangle(|0\rangle+|1\rangle)$ is created by a $\pi/2$ pulse on
the carrier transition followed by a $\pi$ pulse on the blue
sideband. During a waiting time of duration $\tau$, the state
evolves into $|D\rangle(|0\rangle+e^{i\phi}|1\rangle)$, where
$\phi$ is a random variable. Finally, the motional superposition
is mapped back to superposition of internal states
$(|S\rangle+e^{i\phi}|D\rangle)|0\rangle$ and probed by a $\pi/2$
pulse on the carrier transition followed by a quantum state
measurement. We measure the coherence by varying the phase of the
last $\pi/2$ pulse from 0 to $2\pi$ and measuring the contrast
$C(\tau)$ of the resulting Ramsey pattern as a function of $\tau$.
The measurement shown in Fig.~\ref{fig:RamseyContrast} exhibits a
strong reduction in contrast after only 2 ms. This fast decay
could be attributed neither to motional heating (because the
stretch mode heating rate was less than 0.02 phonons/ms), nor to
an instability of the trapping potential (since a coherence time
of about 50~ms was measured on the center-of-mass mode).
\begin{figure}[t]
\includegraphics[width=8cm]{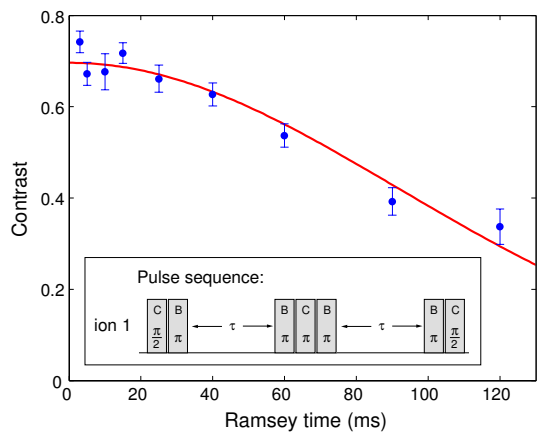}
\caption{\label{fig:SpinechoContrast} Contrast $C(\tau)$ of a spin
echo experiment on the stretch mode as a function of the Ramsey
waiting time $\tau$. Insertion of the spin echo increased the
coherence time by nearly two orders of magnitude as compared with
a simple Ramsey experiment.}
\end{figure}
To probe the dynamics of the dephasing mechanism, a spin echo
experiment was performed on a motional superposition
$|0\rangle+|1\rangle$. For this, the populations of the
$|0\rangle$ and $|1\rangle$ states were exchanged in the middle of
the experiment by a carrier $\pi$-pulse sandwiched between two
blue sideband $\pi$-pulses. As can be seen in
Fig.~\ref{fig:SpinechoContrast}, now it takes the spin echo
contrast $C(\tau)$ about 100 ms to decay to 50\% of its initial
value - a clear indication that the stretch mode frequency is
fairly stable over the duration of a single experiment but
randomly changing from experiment to experiment. This behaviour is
expected for a dephasing caused by a thermally distributed rocking
mode phonon number that takes on random values at the start of
each experiment. To confirm this dephasing mechanism, we
remeasured the Ramsey contrast $C(\tau)$ with both rocking modes
cooled close to the ground state and observed an increase of the
coherence time by a factor of 20 as compared to the experiment
shown in Fig.~\ref{fig:RamseyContrast}.
\begin{figure}[t]
\includegraphics[width=8cm]{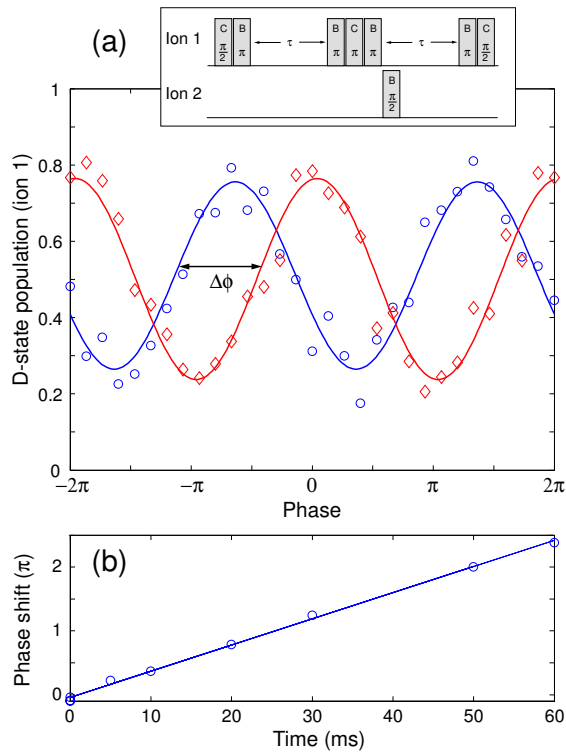}
\caption{\label{fig:Phasescan} (a) Shift $\Delta\phi$ of the spin
echo phase pattern by a rocking mode phonon observed by measuring
the D state population of ion 1 as function of the phase of the
last spin echo pulse. The symbol ($\diamond$) labels experiments
with an extra phonon in the second spin echo time, ($\circ$)
experiments with no extra phonon. (b) Single phonon shift of the
spin echo phase pattern as a function of the waiting time. The
frequency shift caused by a single rocking mode phonon is inferred
from the slope of the function $\Delta\phi(\tau)$.}
\end{figure}

Having established the cross-coupling between the stretch mode and
the rocking mode as the dephasing mechanism, we are interested in
quantifying the cross-coupling strength by measuring the stretch
mode frequency shift caused by a single rocking mode phonon. In
principle, such a measurement could be performed by preparing the
rocking modes in a $n_r=0$ or $n_r=1$ Fock state and subsequently
measuring the stretch mode oscillation frequency in a Ramsey
experiment. This, however, would require ultra-stable high-voltage
sources for keeping the axial trap frequency $\omega_z$ stable to
within $10^{-6}\,\omega_z$ or less. To make the experiment more
robust against technical noise, we carried out a spin echo
experiment instead, where the other ion was excited on the blue
sideband of one of the rocking modes directly before the start of
the second spin echo waiting time $\tau$ (see inset of
Fig.~\ref{fig:Phasescan}(a)). If this pulse excites the second ion
into the $D_{5/2}$ state, there is exactly one additional phonon
in the rocking mode during the second half of the spin echo whose
frequency shift is not compensated by the echo sequence. We adjust
the duration of this blue sideband pulse so that the excitation is
successful in 50\% of the experiments and sort the spin echo data
into two classes according to the quantum state of the second ion
at the end of the experiment. By measuring the phase shift
$\Delta\phi(\tau)$ between the two data sets, we are able to infer
the shift caused by a single rocking mode phonon. This procedure
does not even require cooling the rocking modes to the ground
state. Fig.~\ref{fig:Phasescan}(a) shows the resulting phase shift
for a waiting time $\tau=30$~ms. For a calculation of the single
phonon frequency shift $d\nu_s/dn_r$, we plot the phase shift as a
function of the spin echo time $\tau$, fit a straight line to the
data and determine its slope. For the data measured at
$\omega_z/2\pi=1716$~kHz shown in Fig.~\ref{fig:Phasescan}(b), the
frequency shift is $d\nu_s/dn_r=20.5$ Hz/phonon. This measurement
procedure was carried out for different axial center-of-mass
voltages $\omega_z$ while keeping the transverse oscillation
frequency $\omega_\perp$ fixed.
\begin{figure}[t]
\includegraphics[width=8cm]{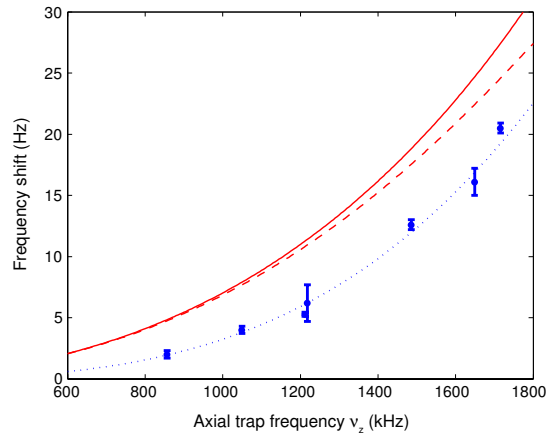}
\caption{\label{fig:StretchmodeVsFrequency} Shift of the stretch
mode frequency pattern by a single rocking mode phonon as a
function of the axial trap frequency $\omega_z$. The solid line is
the frequency shift predicted by eq.~(\ref{freqshift}), the dashed
line shows only the quartic contribution, and the dotted line is a
fit to the data using a heuristic scaling law
$d\nu_s/dn_r\propto\omega_z^\beta$ with $\beta=3.25$.}
\end{figure}
\begin{figure*}
\includegraphics[width=15.5cm]{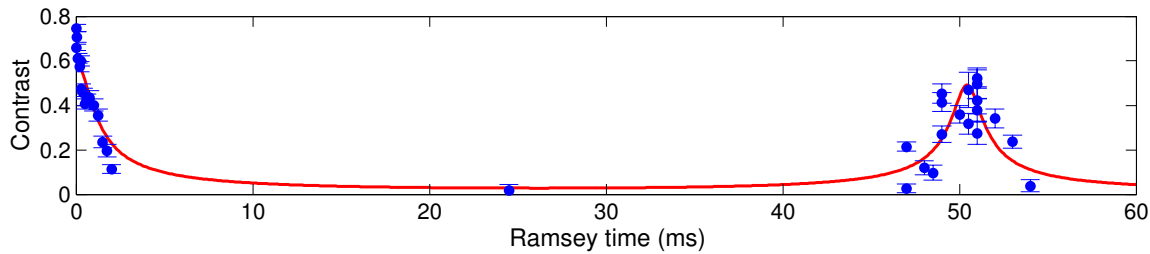}
\caption{\label{fig:StretchmodeRevival} Collapse and revival of
the contrast in a Ramsey experiment probing the motional coherence
of the two lowest stretch mode quantum states. The revival occurs
at the time predicted by the spin echo experiment for a frequency
$\omega_z/2\pi=1716$~kHz.}
\end{figure*}
In Fig.~\ref{fig:StretchmodeVsFrequency} the frequency shift is
plotted as a function of $\omega_z$. The experimentally measured
frequency shift is slightly smaller than the shift predicted by
perturbation theory. The disagreement between experimental data
and theoretical model is currently not understood. It cannot be
attributed to the dynamical nature of the trap potential. Using a
refined pseudopotential \cite{Moore94} leads only to marginal
corrections.

For a thermally occupied rocking mode, the change induced in
$\omega_s$ is always an integer multiple of the single phonon
shift. Therefore, after the initial collapse of the Ramsey
contrast (Fig.~\ref{fig:RamseyContrast}) a revival \cite{Eberly80}
is to be expected for a time
$\tau^\ast=2\pi/\chi=2\pi(d\omega_s/dn_r)^{-1}$. Operating the
trap at $\omega_z /2\pi=1716$~kHz, we indeed observed a revival at
the predicted time $\tau^\ast$. In this experiment, one of the
rocking modes was cooled to the ground state while the other one
was prepared in a thermal state with $\bar{n}_r\approx 9$
(1$\sigma$-confidence interval \{5,17\}).
Fig.~\ref{fig:StretchmodeRevival} shows a fit to the data using
the function
\begin{equation}
\tilde{C}(\tau)=e^{-\gamma\tau}|\langle
e^{i\chi\hat{n}_r\tau}\rangle|\nonumber=e^{-\gamma\tau}|\bar{n}_r+1-\bar{n}_re^{i\chi\tau}|^{-1},\nonumber
\end{equation}
assuming a thermally occupied rocking mode and an overall loss of
contrast $\propto e^{-\gamma\tau}$ accounting for technical noise
and motional heating. A fit to the data yields a revival time
$\tau^\ast=50.5(5)$ ms, an average vibrational quantum number
$\bar{n}_r=9\,(2)$ that is consistent with an independent
measurement of $\bar{n}_r$ and a decay rate
$\gamma=0.004(3)$s$^{-1}$. The existence of revivals is a further
proof of the quantized nature of the rocking motion. The small
loss of contrast shows that the probability of a change in the
rocking mode phonon number within the interval $[0,\tau^\ast]$ is
quite low.

The observed frequency shifts need to be taken into account in
quantum gate realizations operating on the stretch mode. For the
parameters \cite{Leibfried07} used in ref. \cite{Leibfried03}, eq.
(\ref{freqshift}) predicts shifts as big as 100 Hz/phonon, giving
rise to a loss of fidelity of about $0.1\%$ for $\bar{n}=1$.

In summary, we have investigated a nonlinear quantum effect giving
rise to a cross-coupling of harmonic oscillators that can be
described by a Kerr-like Hamiltonian $H\propto\hat{n}_s\hat{n}_r$.
In a two-ion crystal, the nonlinearities lead to a dephasing of
the relative ion motion that manifests itself as a collapse of the
Ramsey contrast that revives once all oscillations get in phase
again. While the nonlinearity is fairly small for the trap
parameters investigated here, it could be made bigger by tuning
the normal mode frequencies closer to the resonance
$\omega_r=2\omega_s$ so that it might be used for creating
entangled motional states of these oscillators. On the other hand,
for quantum computing experiments aiming at fault-tolerant quantum
gates, the observed nonlinearity points to the necessity of
cooling all spectator modes to the ground state or working with
transversally very stiff ion traps.

We acknowledge support by the Austrian Science Fund (FWF), the
European Commission (SCALA, CONQUEST networks), the US Army
Research Office, NSERC and by the Institut f\"ur
Quanteninformation GmbH. C.~R. thanks D.~Leibfried for useful
discussions and F. Dubin for a critical reading of the manuscript.


\begin{thebibliography}{0}
\bibitem{Imoto85} N.~Imoto, H.~A.~Haus, and Y.~Yamamoto, Phys. Rev. A \textbf{32}, 2287 (1985).

\bibitem{Milburn89} G.~J.~Milburn, Phys. Rev. Lett. \textbf{62}, 2124 (1989).

\bibitem{Chuang95} I.~L.~Chuang and Y.~Yamamoto, Phys. Rev. A \textbf{52}, 3489 (1995).

\bibitem{Lloyd99} S.~Lloyd and S.~L.~Braunstein, Phys. Rev. Lett. \textbf{82}, 1784 (1999).

\bibitem{Munro05} W.~J.~Munro, K.~Nemoto, and T.~P.~Spiller, New. J. Phys. \textbf{7}, 137 (2005).

\bibitem{Barrett05} S.~Barrett {\it et al.},
Phys. Rev. A \textbf{71}, 060302 (2005).

\bibitem{Turchette95} Q.~A.~Turchette, C.~J.~Hood, W.~Lange, H.~Mabuchi, and H.~J.~Kimble, Phys. Rev. Lett. \textbf{75}, 4710 (1995).

\bibitem{Schmidt96} H.~Schmidt and A.~Imamo\v{g}lu, Opt. Lett. \textbf{21}, 1936 (1996).

\bibitem{Leibfried03} D.~Leibfried {\it et al.}, Nature \textbf{422}, 412 (2003).

\bibitem{James98} D.~F.~V.~James, Appl. Phys. B \textbf{66}, 181 (1998).

\bibitem{Hoffnagle88} J. Hoffnagle, R.~G.~DeVoe, L.~Reyna, and R.~G.~Brewer, Phys. Rev. Lett \textbf{61}, 255 (1988).

\bibitem{Bluemel88} R.~Bl\"umel {\it et al.}, Nature \textbf{334}, 309 (1988).

\bibitem{Marquet03} C.~Marquet, F.~Schmidt-Kaler, D.~F.~V.~James, Appl.~Phys.~B \textbf{76}, 199 (2003).

\bibitem{King98} B.~E.~King {\it et al.}, Phys. Rev. Lett. \textbf{81}, 1525 (1998).



\bibitem{Roos99} C.~F.~Roos {\it et al.}, Phys. Rev. Lett. \textbf{83}, 4713 (1999).

\bibitem{Schmidtkaler03} F.~Schmidt-Kaler {\it et al.}, J. Phys. B: At. Mol. Opt. Phys. \textbf{36}, 623 (2003).

\bibitem{Schmidtkaler03b} F.~Schmidt-Kaler {\it et al.}, Appl. Phys. B \textbf{77}, 789 (2003).

\bibitem{Moore94} M.~G.~Moore and R.~Bl\"umel, Phys. Rev. A \textbf{50}, R4453 (1994).

\bibitem{Eberly80} J.~H.~Eberly, N.~B.~Narozhny, and J.~J.~Sanchez-Mondragon, Phys. Rev. Lett. \textbf{44}, 1323 (1980).


\bibitem{Leibfried07} D.~Leibfried, private communication.
\end{thebibliography}

\end{document}